\documentclass[11pt]{article}
\usepackage{dcolumn}
\usepackage{multicol}
\usepackage{bm}
\usepackage{amssymb}
\usepackage{amsmath}
\usepackage{cite}
\usepackage{subfigure}
\usepackage[pdftex]{graphicx}
\usepackage[top=.75in,right=.75in,left=.75in,bottom=.75in]{geometry}
\usepackage[left]{lineno}
\usepackage{hyperref}
\usepackage[capitalize]{cleveref}
\usepackage{todonotes}

\begin{document}

\title{Experimental prospects for $V_{ud}$, $V_{us}$, $V_{cd}$, $V_{cs}$ and (semi-)leptonic decays at LHCb}
\author{Adam C. S. Davis\\on behalf of the LHCb Collaboration\\School of Physics and Astronomy\\University of Manchester\\Manchester, United Kingdom}
\date{January 15, 2019}
\maketitle
\begin{flushleft}
\begin{abstract}
	We review the prospects of measurements of the CKM matrix elements corresponding to the first two generations
	of quarks, as well as measurements of leptonic and semileptonic charm meson decays at LHCb. We review the status of searches for 
	Lepton Non-Universality in the charm sector, and provide motivation to continue such searches.\\
	Presented at the 10th workshop on the CKM unitary triangle, Heidelberg, Germany
\end{abstract}
\section{Introduction}
\hspace{5mm}The LHCb experiment at CERN is a single arm forward spectrometer specifically built for the reconstruction of heavy flavor hadron decays
at the LHC\cite{Alves:2008zz}. Given the ample cross-section\cite{xsec}, LHCb has collected the largest sample of charmed and beauty hadrons to date in the world. Leveraging this huge dataset for measurements of the parameters of the CKM triangle has been one of the major programs of the LHCb detector, especially involving bottom quarks, but the prospects for measurements with charm and strange quarks remain relatively unexplored. In these proceedings, we summarize the possible prospects of such measurements at LHCb. We begin with a short review of the standard ways to measure the elements of the CKM matrix involving only the first two generations. Second, we highlight some of the challenges and opportunities of reconstructing semileptonic decays at LHCb and in a hadron collider environment. Next, we offer a few sample measurements in the charm sector. We also review the state of lepton non-universality (LNU) measurements in the charm sector. Finally, we present preliminary sensitivities of such measurements by the LHCb experiment.

\section{The landscape of $V_{qq'}$ measurements} 
\hspace{5mm}Most measurements of the first two generations of the CKM matrix either rely on either fully leptonic decay, semileptonic decay or other processes involving, for instance, nuclear decays\cite{PDG2018}. Precision measurements contributing to the world average of $|V_{ud}|=0.97420\pm0.00021$ currently come from pure vector transitions of nuclei involving $\beta$ decay, resulting in a relative precision of 0.2 per mille, with the limiting uncertainty coming from theory. Measurements using the decay $\pi^+\to\pi^0 e^+\nu_e$ are 30 times less precise, but still in agreement with nuclear $\beta$ decays. Only the pion decay would be possible at LHCb, but these single prong decays are not feasible in the LHCb environment, where charged pions are effectively stable and single kinks are not well reconstructed in the tracking system.\\

\hspace{5mm}Measurements of $|V_{us}|$ are dominated by semileptonic kaon decays, mainly tapping $K_L^0$ and $K^+$ decays, and using the ratio with pion decays to extract $|V_{us}|$\cite{Antonelli:2010yf,PDG2018}. The most general way to report results of semileptonic decays are then to list $|V_{us}|\times f_+(q^2=0)$, then use either lattice information or otherwise to extract $|V_{us}|$ for this decay. Additional information from $\frac{|V_{us}|^2}{|V_{ud}|^2}\frac{f_K^2}{f_\pi^2}$ and inputting the kaon and pion decay constants from leptonic $K$ and $\pi$ decays are not as precise, but still contribute to the world average. The relative precision for $|V_{us}| = 0.2243\pm 0.0005$\cite{PDG2018} now stands at 2 per mille. \\

\hspace{5mm}In the LHCb Upgrade II, given the extremely high cross-section of strange hadrons, measurements of branching fractions at the level of $10^{-15}$ are possible, assuming perfect reconstruction efficiencies\cite{Junior:2018odx}. This opens the door to the possible first observation of the $K_s^0\to\pi^+\mu^-\nu_\mu$ decay, and possible contribution to the measurement of $|V_{us}|f^+(q^2)$. Additional measurements are possible utilizing the decay $\Lambda^0\to p\mu^+\nu_\mu$, but in both cases, lifetime of the $K_S^0$ and the $\Lambda$ will have significant impact on both the ability to trigger these candidates as well as their reconstruction, given that most
will decay downstream of the vertex locator, increasing the difficulty of vertexing.

\hspace{5mm}Decays of charmed mesons allow for the extraction of $|V_{cd}|$. The usual methods involve either the purely leptonic decay
$D^+\to \ell^+\nu_\ell$, or semileptonic decays of a singly charmed hadron $H_c\to H_d \ell \nu_\ell$. Current world averages give $|V_{cd}| = 0.218\pm 0.004$, with a relative precision of about 2\%\cite{PDG2018}. Previous measurements include neutrino scattering, a technique which is beyond the capabilities of the LHC.
Purely leptonic single prong $D^+$ decays are also not favored, as in the case of the pion decay. However, the decays $D^0\to\pi^-\ell^+\nu_\ell$ are accessible at LHCb, especially if originating from the decay chain $D^*(2010)^+\to D^0 \pi_s^+$. As these are abundantly produced, the main uncertainties will be with the lepton chosen, as well as the reconstruction method of the missing neutrino. We will approach how to deal with these uncertainties in the coming sections. Further decays involving $\Lambda_c$ baryons are possible, though much more challenging, as the lifetime of the $\Lambda_c^+$ baryon is a factor two less than that of the $D^0$, which limits prompt production trigger efficiency. Tagging from secondary $b$-hadron decays is possible, but suffers from lower statistics due to the difference in cross-sections between $c-$ and $b-$hadrons.

\hspace{5mm}The extraction of $|V_{cs}|$ is often done in analogy to that of $|V_{cd}|$, either substituting a kaon for a pion, or $D_s^+$ for the $D^+$ decays. As is the case with $|V_{cd}|$, world averages of $|V_{cs}|=0.997\pm0.017$, places the relative precision just below 2\% \cite{PDG2018}, and as with the $D^+$ decays, single prong $D_s^+$ decays remain out of reach of LHCb measurements. Measurements involving on-shell $W$ decays is out of scope of LHCb analyses. Semileptonic decays, however, are still fair game, and the Cabibbo favored $D^0\to K^-\ell^+\nu_\ell$ decays reconstructed at LHCb will easily have the largest statistics of any sample in the world. The measurements then rely on the explicit cancellation of detector and theoretical uncertainties, which motivates the measurement of $\frac{|V_{cd}|\times f_+^{D^0\to\pi}(q^2)}{|V_{cs}|f_+^{D^0\to K}(q^2)}$ using $R=\frac{\mathcal{B}(D^0\to\pi\mu\nu)}{\mathcal{B}(D^0\to K\mu\nu)}$. Utilizing this ratio, and measuring as a function of $q^2$ leaves only the efficiencies kaons and pions to be calculated, which are well known at LHCb\cite{Alves:2008zz,LHCb-PROC-2011-008}.

\hspace{5mm}An interesting new approach is to extract of $|V_{cd}|$ and $|V_{cs}|$ using the reconstruction of the decays $B_c^+\to B^0_{(s)}\mu^+\nu_\mu$ and $B_c^+\to B^0_{(s)}\pi^+$, tapping into the rich $B_c$ physics program of LHCb\cite{LHCbPhaseII}. We propose these measurements here for the first time. For extraction of only $|V_{cs}|$, one can also look to measure $B_c^+\to B^0 K^+$. As with the semileptonic measurements of $D^0\to K^-\ell^+\nu_\ell$ and $D^0\to\pi^-\ell^+\nu_\ell$, taking the ratio of different channels allows for the effective cancellation of many uncertainties, and leaves
more theoretically clean observables. Some limitation are high statistics samples of $B_c$ mesons, and the knowledge of the $B_c$ hadronization fractions. Both of these points will be improved by the LHCb experiment in years to come via direct measurements, but it may also be possible to avoid some of these constraints by instead of focusing on the CKM elements, taking ratios with the $D$ sector to isolate the form factors and hadronization fractions.

\hspace{5mm}For the remainder of these proceedings, we will focus on the measurements involving semileptonic charm meson decays to extract quantities of interest.

\section{Reconstruction of semileptonic decays at LHCb}
\hspace{5mm}The hadron collider environment of LHCb does not allow for the conventional techniques of reconstruction employed at $e^+e^-$ colliders; one cannot simply rely on the beam energies and full reconstruction of a tag side of the decay to solve for the momentum of the missing neutrino. Partial reconstruction is already known given a displaced secondary vertex, namely the momentum of the neutrino perpendicular to this flight direction. The remaining component has a two fold ambiguity with respect to this flight direction. One can either choose to not reconstruct this component or reconstruct given extra information. If one chooses not to reconstruct the component, then can use the so-called ``corrected mass'' $m_{corr} = \sqrt{p_T'^2 + m_{vis}^2} + |p_T'|$; using the example of the decay $D^0\to K^- \mu^+ \nu_\mu$, the variable $p_T'$ is the momentum component of the visible system perpendicular to the $D^0$ flight direction, and $m_{vis}$ is the visible invariant mass of the fully reconstructed candidates, in this case $m(K\mu)$. This variable provides distinguishing power, as if there is truly one massless particle missing, then it will peak at the correct mass, and have a tail to lower values, whereas if there are more missing particles, the distribution will be shifted to lower values and not peak as strongly. This is show in Figure~\ref{fig:corrm}, generated using RapidSim\cite{rapidsim}. Other reconstruction methods include the well-known $k-$factor approach, where the true momentum of the $D^0$ candidate is given with a function of the visible mass and a factor $k$ to solve for the visible momentum, $p(K\mu\nu)=\frac{p(K\mu)}{k(m_{vis})}$, multivariate regression\cite{Ciezarek:2016lqu}, and using higher resonance decays to break the two-fold ambiguity of momentum by adding an additional mass constraint, for instance using the cone closure method\cite{coneclosure}. One can also use multiple methods to exploit differing correlations between signal and background across multiple variables.

\begin{figure}[htbp]
\begin{center}
\includegraphics[width=0.5\textwidth]{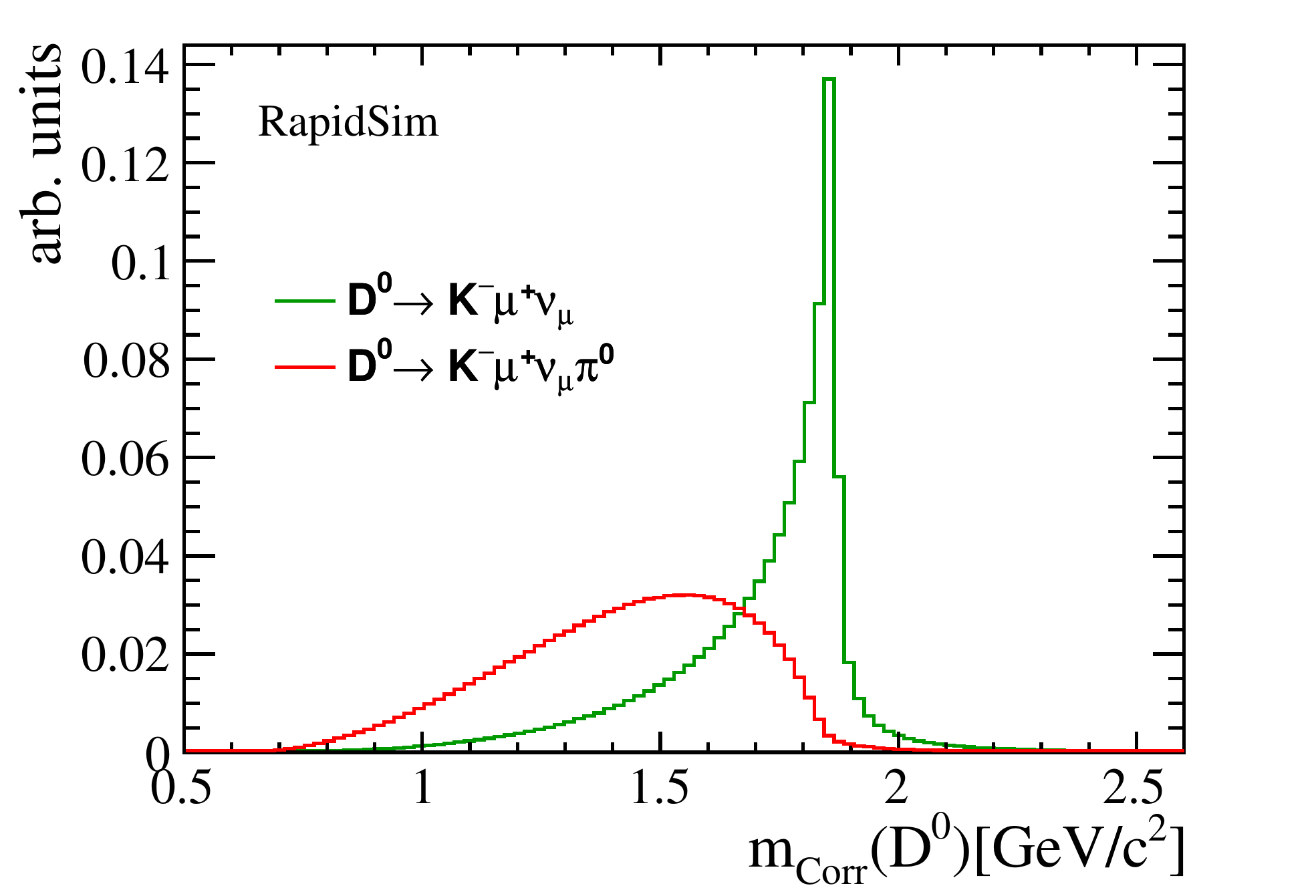}
\caption{Corrected mass distributions for $D^0\to K^-\mu^+\nu_\mu$ (green) and $D^0\to K^-\mu^+\nu_\mu\pi^0$ (red) decays. Uses RapidSim\cite{rapidsim}.}
\label{fig:corrm}
\end{center}
\end{figure}

\section{Sensitivity estimates for $\frac{|V_{cd}|f_+^{D^0\to \pi}(q^2)}{|V_{cs}|f_+^{D^0\to K}(q^2)}$ at LHCb}
\hspace{5mm}The differential decay rate of the decay $D^0\to P^-\ell^+\nu_\ell$ at leading order can be written as\cite{Aoki:2016frl}
\begin{equation}
\label{eq1}
\begin{split}
	\frac{d\Gamma(D^0\to P^- \ell^+ \nu_\ell)}{dq^2} = |V_{cQ}|^2 & \frac{G_F^2}{24\pi^3}\frac{(q^2-m_\ell^2)^2\sqrt{E_P^2-m_P^2}}{q^4m_{D^0}^2}\\ &\times \left[\left(1+\frac{m_\ell^2}{2q^2}\right)m_{D^0} (E_P^2-m_{P}^2)|f_+(q^2)|^2 + \frac{3m_\ell^2}{8q^2}(m_{D^0}^2-m_P^2)^2 |f_0(q^2)|^2\right],
\end{split}
\end{equation}
where $Q$ represents the outgoing quark from the weak vertex, the terms $f_+(q^2)$  and $f_0(q^2)$ are the vector and scalar form factors as a function of the momentum transfer squared to the lepton-neutrino pair $q^2$, respectively, used to parameterize the hadronic current. As the scalar form factor is suppressed by the factor $m_\ell^2/m_{D^0}^2$, it is customary to omit this form factor, leading to the simplified form
\begin{equation}
	\frac{d\Gamma(D^0\to P_x \ell \nu)}{dq^2} =\frac{G_F}{24\pi^3}\left| \vec{p}_P\right|^3 \left|V_{cx}\right|^2 |f_+(q^2)|^2.
\end{equation}
Therefore, taking the ratio of the differential decay rates of the decays $D^0\to \pi^-\mu^+\nu_\mu$ and $D^0\to K^-\mu^+\nu_\mu$ results in a measurement of $\frac{|V_{cd}|\times f_+^{D\to\pi}(q^2)}{|V_{cs}|\times f_+^{D\to K}(q^2)}$. By utilizing the cone closure technique from $D^{*+}\to D^0\pi^+$ decays, one can accurately reconstruct the $q^2$ distributions of the $D^0$ decay. Preliminary yields of the $D^0\to K^-\mu^+\nu_\mu$ decays from this channel give $q^2$ integrated yields of roughly 5 million signal candidates for Run~1 of the LHC\cite{asls,Davis:2017ckr}, which would yield a relative uncertainty on the measurement of 0.2\%. If the error on the form factors can at some point be neglected, this would result in a measurement of the CKM elements at the level of the current world average\cite{PDG2018}.

\section{Sensitivity of measurements of Charge Parity Violation and mixing from $D^0\to K\mu \nu_\mu$}
\hspace{5mm}The decay $D^0\to K\mu\nu$ is interesting itself from a mixing standpoint as well: the right-sign $D^0\to K^- \mu^+ \nu_\mu$ is entirely dominated by Cabibbo-favored decay, whereas the wrong-sign decay is only accessible from mixing. This leaves the ratio of wrong-sign to right-sign decays as
\begin{equation}
	R(t)=\frac{\mathcal{P}(D^0\to K^+\mu^-\overline{\nu}_\mu)}{\mathcal{P}(D^0\to K^- \mu^+ \nu_\mu)} \propto \frac{x^2+y^2}{4}\left(\frac{t}{\tau}\right)^2,
\end{equation}
where $\mathcal{P}$ represents the probability of the transition, $x= 2(m_1 -m_2)/(\Gamma_1+\Gamma_2)$, y = $(\Gamma_1-\Gamma_2)/(\Gamma_1+\Gamma_2)$, and $t$ is the $D^0$ decay time. Here $m_{1,2}$ is the mass of the different $D$ mass eigenstates, and $\Gamma_{1,2}$ is the width of the different eigenstates.\\
\hspace{5mm}Integrating over decay-time, this ratio is $R_M=\frac{x^2+y^2}{2}$. Using the same 5M as a sample as a baseline and the current world averages for $x$ and $y$, the projected number of WS decays would be around 700, corresponding to an uncertainty on $R_M$ of the order of 0.01\%\cite{dmitzel}. These projections, however, do not include form-factor uncertainties, nor shape uncertainties in the fit. Extraction of CPV related quantities would be possible by splitting by the charge of the pion from $D^{*+}\to D^0 \pi^+$ decays.

\section{Lepton non-universality in charm}
\hspace{5mm}As the flavor anomalies in the $b$-hadron sector still persist, it becomes increasingly important to search in different regions for similar effects. One such place is the semileptonic decays $D^0\to P\ell\nu_\ell$. As is evident from equation~\ref{eq1}, taking the ratio between electron and muon modes will lead to large cancellations in form factor uncertainties, leaving a theoretically clean ratio. This represents a measurement of lepton non-universality in charm decays (LNU). Measurements of these types have only now started to be completed by the BESIII collaboration\cite{Ablikim:2018evp,Ablikim:2018frk}, and LHCb measurements are expected soon.

\hspace{5mm}The state of global measurements is shown in Figure~\ref{fig:lnu}. As pointed out in the previous version of this plot\cite{Davis:2017ckr}, all points lie to one side of this line. Current progress from BESIII maintains this trend, but increases the precision.

\begin{figure}[htbp]
\begin{center}
\includegraphics[width=0.6\textwidth]{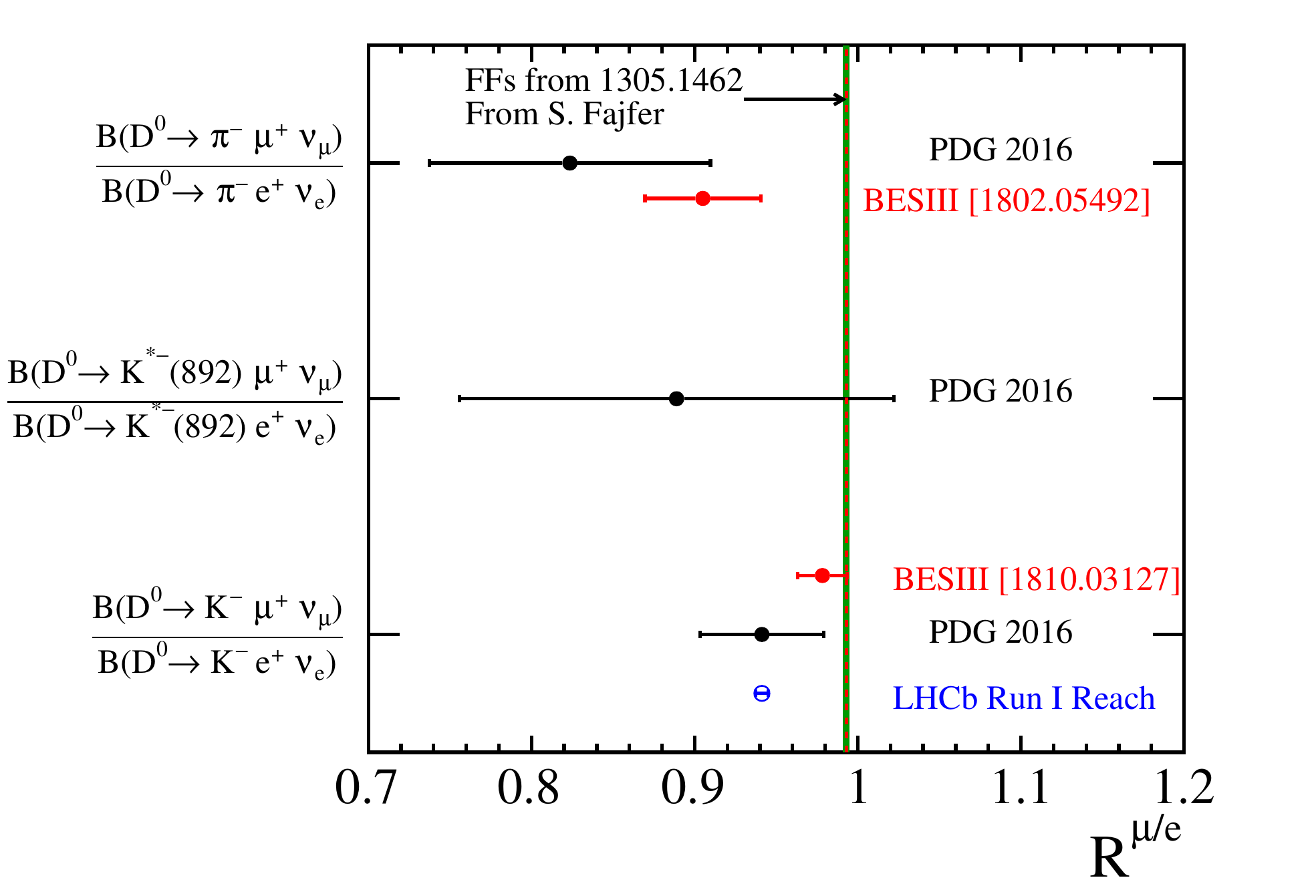}
\caption{Summary of measurements of lepton non-universality in $D^0\to P\ell\nu$ decays. Reported is the ratio of muon ($D^0\to P^- \mu^+\nu_\mu$) to electron ($D^0\to P^- e^+\nu_e$) decays for $P=K^-,\pi^-,K^*(892)^-$. In black are the ratio of muon to electron branching fractions as reported in the PDG 2016 edition \cite{PDG2016}, in red are the current BESIII measurements, and in blue the estimated reach of LHCb from the Run~1 dataset, with the central value set to the PDG 2016 value. The bottom-most BESIII measurement is also provided in Ref.~\cite{Ablikim:2018evp}. The theoretical prediction given by S. Fajfer is shown in the red dashed line with the one sigma uncertainty listed as the green hatched area.}
\label{fig:lnu}
\end{center}
\end{figure}

\hspace{5mm}Given the dataset from LHCb, and incorporating differing efficiencies for different triggering, the Run~1 reach is estimated to increase the precision from the 2016 average by one order of magnitude on the $q^2$ integrated measurement. In the case of $D^0\to K^-\ell^+\nu_\ell$ decays, possible new physics effects are also probed in the $q^2$ dependence of the LNU ratio $R^{\mu/e}$ which are not present in the measurement of the branching ratio alone, due mainly to form factor uncertainties\cite{Fajfer:2015ixa}. The BESIII collaboration has recently measured the $q^2$ dependence of this ratio, but lacks the precision to distinguish between standard model and new physics effects. The LHCb collaboration will provide the measurement of $R^{\mu/e}$ vs $q^2$ by utilizing the $D^{*+}\to D^0\pi^+$ decay, and using the cone closure technique. A pseudo-experiment fit is shown for illustrative purposes of one of the two fit methods considered in Figure~\ref{fig:lnufit}. The method is a two-dimensional fit to the corrected mass of the $D^0$ meson versus the visible mass difference $\Delta m_{visible} = m(K\ell\pi_s)-m(K\ell)$. The different components of peaking backgrounds and signal are shown in different colors.

\begin{figure}[htbp]
\begin{center}
\includegraphics[width=0.6\textwidth]{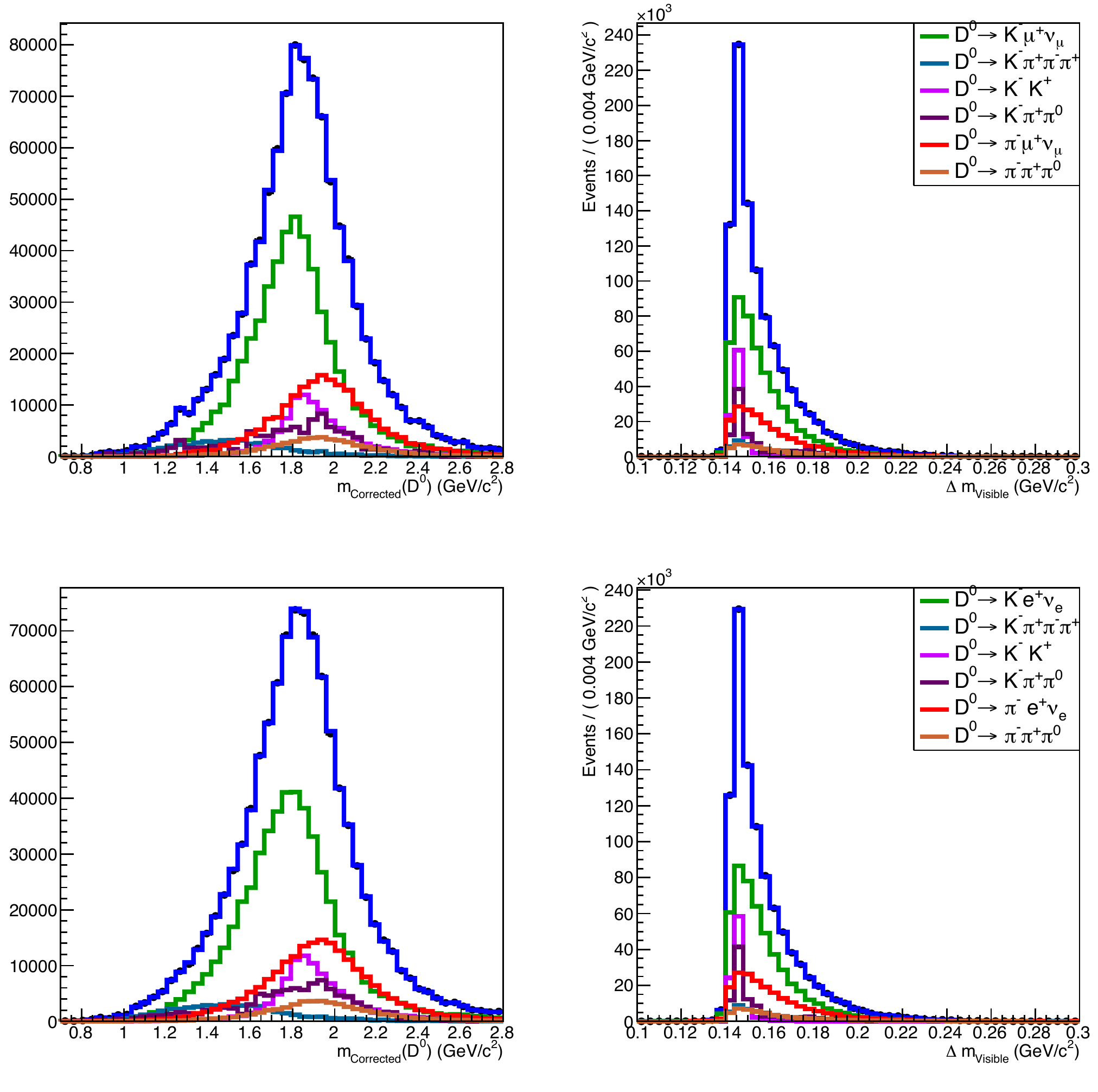}
\caption{Unofficial fit to one pseudo-experiment of $D^0\to K^-\mu^+\nu_\mu$ (top) and $D^0\to K^-e^+\nu_e$ (bottom) decays at LHCb. The fits show are projections onto the corrected $D^0$ mass and the visible delta mass, left and right, respectively. Individual fit components are shown in different colors.}
\label{fig:lnufit}
\end{center}
\end{figure}

\section{Conclusions}
\hspace{5mm}LHCb has collected the largest charm sample in the world, and despite operating in a hadron collider environment, offers strong opportunities to increase knowledge in CKM elements in the first two generations, provide sensitivity to mixing and CPV in the $D^0$ system, and unique contributions to LNU searches in the charm sector utilizing semileptonic decays. Many sensitivities were presented for the Run~1 dataset, with Run~2 providing an additional $\sim6~$fb$^{-1}$ at $\sqrt{s}=13$~TeV. Clearly the contribution of the LHCb experiment in this field are both motivated and needed, and contributions will come soon. Additionally, new ideas on how to access quantities of interest, both in rare kaon decays and leveraging the $B_c$ physics program will contribute to the knowledge in the sector.


\end{flushleft}
\bibliographystyle{h-physrev}
\bibliography{CKMProceedingsbib}
\end{document}